\documentclass[twocolumn,showpacs,preprintnumbers,amsmath,amssymb]{revtex4}  

\usepackage{graphicx}
\usepackage{dcolumn}
\usepackage{bm}

\begin{document}

\title{How does the past of a soccer match influence its future? }


\date{\today}

\author{Andreas Heuer}
\affiliation{\frenchspacing Westf\"alische Wilhelms Universit\"at M\"unster, Institut f\"ur physikalische Chemie, Corrensstr.\ 30, 48149 M\"unster, Germany}
\affiliation{\frenchspacing Center of Nonlinear Science CeNoS, Westf\"alische Wilhelms Universit\"at M\"unster, Germany}
\author{Oliver Rubner}
\affiliation{\frenchspacing Westf\"alische Wilhelms Universit\"at
M\"unster, Institut f\"ur physikalische Chemie, Corrensstr.\ 30,
48149 M\"unster, Germany} \affiliation{\frenchspacing Center of
Nonlinear Science CeNoS, Westf\"alische Wilhelms Universit\"at
M\"unster, Germany}


\begin{abstract}

Scoring goals in a soccer match can be interpreted as a
stochastic process. In the most simple description of a soccer
match one assumes that scoring goals can be described by a
constant goal rate for each team, implying simple Poissonian and
Markovian behavior.  Here a general framework for the
identification of deviations from this behavior is presented. For
this endeavor it is essential to formulate an a priori estimate of
the expected number of goals per team in a specific match. The
analysis scheme is applied to approximately 40 seasons of the
German Bundesliga. It is possible to characterize the impact of
the previous course of the match on the present match behavior.
This allows one to identify interesting generic features about
soccer matches and thus to learn about the hidden complexities
behind scoring goals.

\end{abstract}






\maketitle

\section{Introduction}

In recent years researchers from the physics community have
started to apply  physics-oriented analysis to problems from the
area of sports and  in particular of soccer \cite{Wesson, Tolan,heuer_buch}.
Specific examples for a quantitative analysis of the outcome of
sports events can be found, e.g., in
\cite{suter,ben1,ben2,janke1,janke2} and new ranking schemes have been proposed \cite{Radicchi}.  At first one might think
that it is hard to find systematic laws to characterize such
complex phenomena as soccer matches. One key step in this endeavor
is the definition of appropriate observables to capture some key
properties. In recent years we have concentrated on the formal
characterization of the notion of a team strength and its
practical determination \cite{Heuer1}. In this way it was possible
to ask questions about the variation of team strength during a
season \cite{Heuer2} or the impact of a coach dismissal on the
team strength \cite{Heuer3}. Alternative concepts of team strengths have been studied, e.g., in Ref.\cite{Sire} for the case of baseball.

Already a long time it has been realized that the distribution of
goals, scored by a team, can be roughly described by a Poisson
distribution \cite{Mah82, Lee97, Rue00}. Such a distribution is to
be expected if the probability to score a goal in the next minute
is constant within the whole match. In the most simple stochastic
model of a soccer match one might simply assume that both teams
score goals according to {\it independent} Poisson
distributions. Closer inspection of the empirical goal
distribution displays, however, some broadening as compared to a
Poisson distribution.   To rationalize this observation a model
has been presented which postulates an increase of the goal rate
with an increasing lead \cite{janke1,janke2}. This
self-affirmative effect could indeed reproduce the fat tails in
the empirical goal distribution. In later work it has been shown
that at least for the German soccer league (Bundesliga) these fat
tails just follow from the distribution of team strengths
\cite{Heuer2}. Therefore the fat tails do not contradict the
notion that in an individual match the scoring of goals follows
Poisson statistics without self-affirmative effects.

Interestingly, it turns out that the number of draws is
significantly larger (approx. 10\%) than expected from the assumption of
independent Poisson distributions\cite{Rue00}. Different
scenarios may lead to this effect. Here are two extreme cases: (1)
A draw in the, let's say, 70th minute reduces the attempts of both
teams to score another goal. This leads to an increased
probability to keep this score. (2) A score of, e.g., 1:0, may
strongly enhance the willingness of the trailing team to score a
goal to reach at least a draw. Whether or not any of these
scenarios indeed explain the excess of draws is not clear a
priori. Knowledge of such effects would allow one to gain
information about psychological effects within a soccer match. The
central aim of this work is derive a stochastic description of the
course of a soccer matches without resorting to any ad hoc models.
Recently, somewhat related questions have been analysed, e.g.,  for the case of basketball \cite{Gabel} and tennis \cite{Magnus}. These results can then be compared with the present analysis.

The structure of this paper is as follows. In Sect.2 we discuss
the statistical  framework to elucidate the basic complexities of
a soccer match. In Sect.3 the results of this analysis are
presented which are finally discussed in Sect.4.

\section{Statistical framework}

In a specific match of team A vs. team B one may estimate the
number of expected goals $\lambda_A$ of team A and $\lambda_B$ of
team B based on the strength of both teams
\cite{Dixon97,Rue00,Dobson03}. Here we choose the approach as used
in Ref. \cite{Heuer2}. In more detail, by taking the goal difference and the sum of all goals for the 33 other matches of both teams, considering the regression towards the mean, and adding a team-independent home advantage (see Ref.\cite{Heuer1}) one can indeed obtain good estimates of $\lambda_{A,B}$. In what follows we define the goal rate as
the probability to score a goal in the next minute. If the goal
rate of, e.g., team A does not change during the match one can
define the goal rate $\gamma_A$ via $\gamma_A = \lambda_A/90$.
Note that a soccer match lasts for 90 minutes.

\begin{figure}[tb]
\centering\includegraphics[width=0.7\columnwidth]{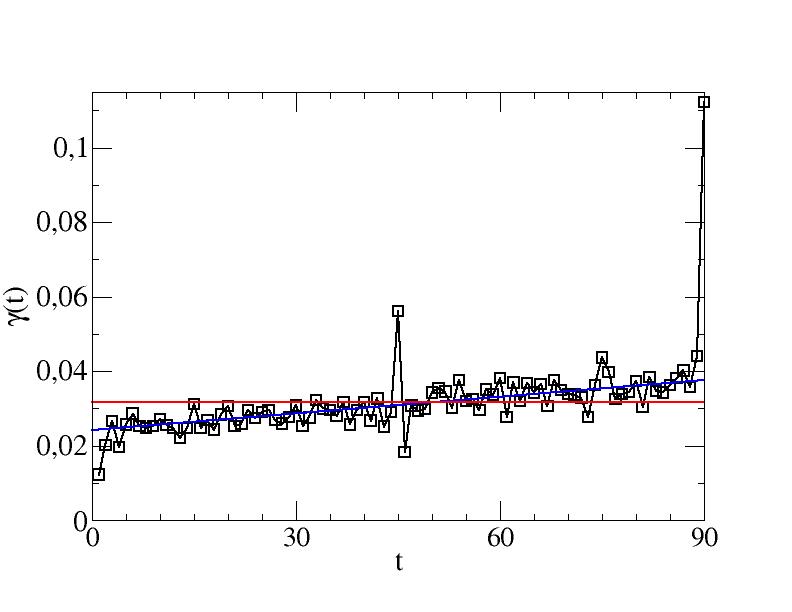}
\caption{The average number of goals per minute in a match as a
function of the time t. A more detailed interpretation of the data can be found in the subsequent section.} \label{fig1}
\end{figure}

It is known that in the second half of a soccer match
significantly more goals are scored than in the first half (43\%
vs. 57\%). Thus, one may expect that typically the goal rate $\gamma_A(t)$
increases with time. To capture this effect more quantitatively, we
introduce $ \gamma_{tot}(t) = \langle \gamma_A(t) + \gamma_B(t) \rangle $ as the
goal rate in minute $t$, averaged over all matches. Note that $\gamma_{tot}(t)$ takes into
account the goals of both teams.

Here we
consider the Premier German soccer league (Bundesliga) between
seasons 1968/69 and 2010/11 (excluding 1992/93 because the number
of teams changed in this particular season). The resulting curve
for $\gamma_{tot}(t)$ is shown in Fig.\ref{fig1}. One can indeed see the
general increase of $\gamma_{tot}(t)$ with time. Some additional
specific features of Fig.\ref{fig1} will be discussed in Sect.
III. When summing up $\gamma_{tot}(t)$ over all 90 minutes one obtains
the total number of goals per match, denoted $\lambda_{tot}$. One
finds $\lambda_{tot}=3.07$. Correspondingly, a single team on
average scores 1.53 goals per match.

For the time being we  assume that
it is indeed possible to define and determine a goal rate $\gamma_A(t)$ of team A at a given minute
in an {\it individual} match. In reality this is impossible because playing soccer
is much more complex than throwing a dice. The arrangement of the soccer players and the ball
at some moment and possibly all other available pieces of information only allows a very rough estimation of the
probability of a goal during the next minute. However, for the results of this work it will be sufficient
to consider averages
over a large number of appropriately selected matches. Therefore, in practice the average goal rate will
be simply determined from counting the matches where a goal was scored in the minute under consideration.

For future purposes we introduce the normalized rate
\begin{equation}
\label{ga}
\Gamma_{A}(t) = \frac{\gamma_{A}(t)}{\gamma_{tot}(t)}.
\end{equation}
We remind again that the nominator contains the rate for an individual match whereas the denominator
expresses the average over all matches. In any event $\Gamma_A(t)$ reflects the real course
of the match. Furthermore we use
the normalized expected number of goals in a specific match of team A
\begin{equation}
\Lambda_A = \frac{\lambda_A}{\lambda_{tot}}.
\end{equation}
In contrast to Eq.\ref{ga} $\Lambda_A$ expresses the a priori expectation.

In general, the function $\gamma_A(t)$ can be very complicated and can vary from match to match.
For the future discussion it is helpful to identify a limit of maximum simplicity which
we denote as the Poisson expectation. It can
be formulated via two conditions:
 (1) The integral of the goal rate $\gamma_A(t)$
over the whole match is identical to the expected number of goals
$\lambda_A$ of team A in this match.  (2) The time-dependence of
$\gamma_A(t)$ is, apart from a proportionality factor, identical
to that of $\gamma_{tot}(t)$ and thus follows the average behavior as shown in Fig.\ref{fig1}.
As a consequence one would have
\begin{equation}
\label{simple}
\gamma_{A}(t) = \frac{\lambda_{A}}{\lambda_{tot}} \gamma_{tot}(t)
\end{equation}
which is equivalent to
\begin{equation}
\label {eqsimp} \Gamma_A(t) = \Lambda_A.
\end{equation}
For this simple Poisson expectation the actual normalized goal rate $\Gamma_A(t)$ for
a specific match thus equals the pre-match expectation $\Lambda_A$.

The key goal of this work is to identify situations where
Eq.\ref{eqsimp} and thus Eq.\ref{simple} are not valid. For
example we consider all matches for which the home team A just
before minute 80 leads by m=3 goals and ask for the probability
that the home team scores a goal in the next minute. In what
follows $m$ always denotes the goal difference. Then we define
$\Gamma_h(t=80,m=3) \equiv \langle \Gamma_A(t=80)
\rangle^\prime_A$. The prime indicates a conditional average. In
this specific case we average the normalized goal rate of the home
team over all matches for which the home team fulfills the
required condition ($m=3$ at minute 80). The actual calculation of
$\Gamma_h(t=80,m=3)$ basically boils down to the calculation of
$\langle \gamma_A(t=80)\rangle^\prime_A$ which, according to our
conditioning, denotes the fraction of matches with $m=3$ at minute
80 for which the home team scores a goal in minute 80. In analogy,
we define $\Lambda_h(t=80,m=3) \equiv \langle \Lambda_A
\rangle^\prime_A$ as the corresponding expectation value of
$\Lambda_A$ for the same subset of teams. Deviations from the
relation $\Gamma_h(t=80,m=3) = \Lambda_h(t=80,m=3)$  and the
corresponding violation of  Eq.\ref{eqsimp} directly imply that
with a home lead of three goals in minute 80 the match behavior is
different as expected from the simple Poisson expectation. As will
be shown in this work for a home lead by three goals the
probability of increasing the home lead is smaller than expected
from the Poisson expectation. We note in passing that without
conditioning, i.e. by averaging over all matches and all teams,
one obtains by definition $\langle \Gamma_A(t) \rangle = \langle
\Lambda_A \rangle = 1/2$.

\begin{figure}[tb]
\centering\includegraphics[width=0.7\columnwidth]{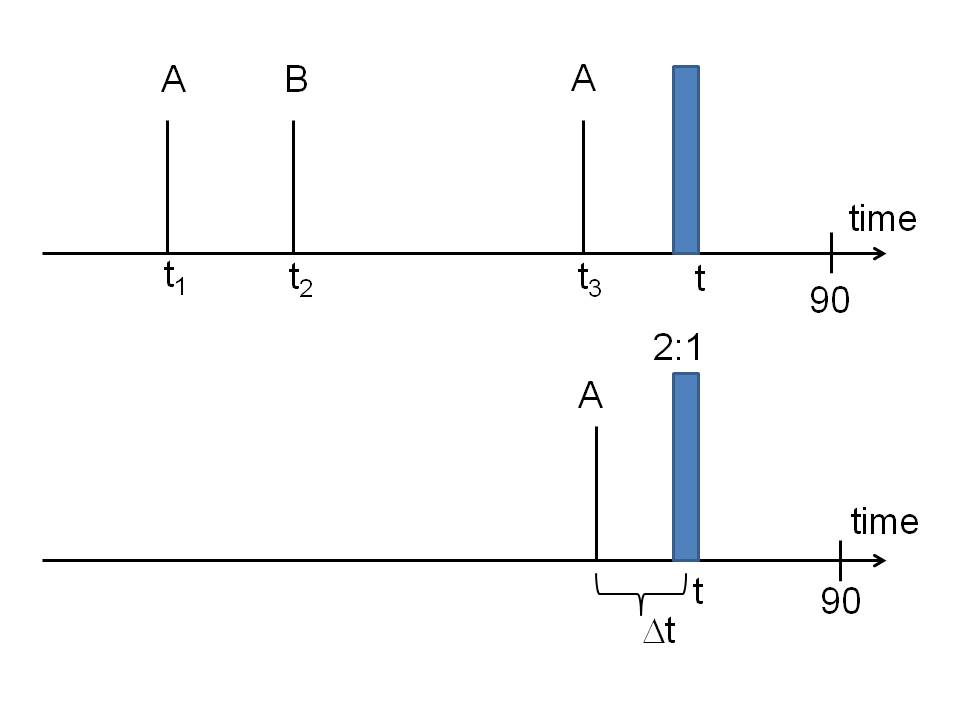}
\caption{Upper row: the complete information about the goals for a
specific example which contribute to the prediction of the goal
rate $\Gamma_A(t)$ at time t. Lower row: the reduced information
which takes into account the score at time t as well as the time
difference $\Delta t$ to the last goal and the information about
the team which scored that goal.} \label{sketch}
\end{figure}

An important first step is the {\it systematic} identification of the most relevant
items for the conditioning of $\Gamma_A(t)$. Let's assume that in minute 70 in total $g=3$ goals have been
scored and that the actual score is 2:1 for team A. The goals of
team A were scored in minutes 10 and 60 and the goal of team B in
minute 25. Strictly speaking we want to understand the impact of
the previous goal events on the goal rate at a given time $t$.
$\Gamma_A(t)$ is thus conditioned on the sequence of the previous
goals as well as the precise time of these goals. This is
illustrated in the upper part of Fig.\ref{sketch}. Apart from the
fact that this complete dependence is impossible to extract from
the available information one can dramatically simplify the
required conditioning. For a specific example it will be shown
that neither the order of the goals (e.g. 0:1 $\rightarrow$ 1:1
vs. 1:0 $\rightarrow$ 1:1)  nor the {\it absolute} times
$t_1,...,t_g$  of the goals play a relevant role. We just mention
in passing that the latter disagrees with the general belief that
goals just before half time are particularly helpful for a team.
These observations strongly suggest that in general the dependence
on the order and the absolute times of the previous goals is, if
existent at all, very weak. Thus, neglecting these pieces of
information does not reduce the estimation quality of
$\Gamma_A(t)$. As a strict consequence $\Gamma_A(t)$  can only
depend on the following observables: (1) Score in minute $t$. What
is the expected course of the match during minute 70  if, e.g.,
the home team leads by one goal? In what follows we mainly
restrict ourselves to the goal difference rather than to the
absolute number of goals. (2) {\it Relative} time differences $t -
t_i$ ($i \in {1,...,g})$. One may indeed expect that scoring a
goal may give rise to a minor shock to the opponent which, as a
consequence, may bias the match during the minutes after the goal.
Naturally the impact of the last goal is strongest. Thus, we only
keep track of the time difference $\Delta t = t - t_g$. This
reduction of information is summarized  in the lower part of
Fig.\ref{sketch}. As soon as the goal rate depends on the present
score one leaves the regime of Poisson processes and, in general,
(possibly small) deviations from a strict Poisson goal distribution would be expected. Furthermore,
any dependence on the time elapsed since the previous goal is a
clear signature of non-Markovian effects since memory effects
start to play a role.

\section{Results}

\subsection{Total number of goals}

We start with the disucssion of the time dependence of the total
number of goals in minute $t$, i.e. $\gamma_{tot}(t)$. The data were
already shown in Fig.\ref{fig1}. Except for some specific minutes
one observes a linear increase of the total goal rate with time.
As compared to this linear trend the values in the 1st and the
46th minute are reduced by approx. 50 \%, respectively. This
expresses the fact that the initial condition of the match (full
separation of both teams in the halves of the field) implies a
minimum time until the first goal can be scored. Roughly speaking,
equilibration is reached after 30 seconds. The spikes in the 45th
and 90th minute have the trivial origin that a match typically has
some overtime which, however, is counted as minute 45 (after the
first half) or as minute 90 (at the end). Somewhat surprisingly,
no deviations from the linear trend are seen after the half time
break (except for the obvious reduction of the goal rate at minute
46). This means that the match is basically continued as if there
had not been any break. Furthermore a significant increase is
observed beyond minute 87, representing an increasing offensive
(or decreasing defensive) behavior.

The time-dependent rate still allows the whole process to be
Poisson. Let us, for reasons of simplicity, consider the case
where the average number of goals in the first half is $\lambda_1$
and in the second half $\lambda_2$. Then the probability to have
$g$ goals in the total match can be written as
\begin{equation}
p(g) = \sum_{g_1}\sum_{g_2} p_{Poisson}(g_1,\lambda_1)
p_{Poisson}(g_1,\lambda_2) \delta_{g,g_1+g_2}
\end{equation}
where $p_{Poisson}(g,\lambda)$ denotes the standard Poisson
distribution. Application of the binomial equation yields after a
straightforward calculation $p(g) =
p_{Poisson}(g,\lambda_1+\lambda_2)$. Thus, a time-dependent goal
rate still allows the match outcome to fulfill Poisson statistics.

\begin{figure}[tb]
\centering\includegraphics[width=0.7\columnwidth]{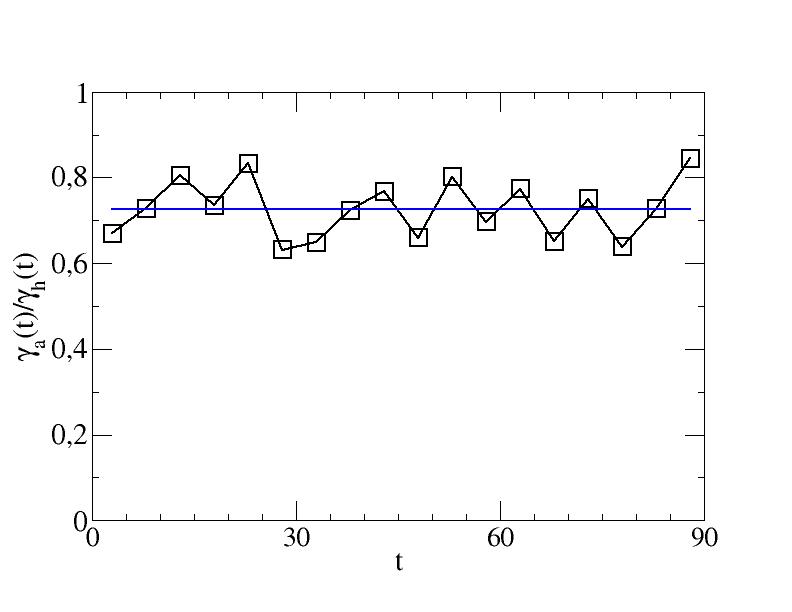}
\caption{The ratio of the average number of goals of the away team
vs. those of the home time as a function of the time t. }
\label{fig2}
\end{figure}

It may be instructive to analyze the ratio  of goals, scored by
the away team and the home team, respectively, as shown in
Fig.\ref{fig2}. It turns out that within fluctuations this ratio
is constant throughout the match. Thus, no additional
home-away-asymmetry has to be taken into account for the
statistical description of the total goal rate.

\subsection{Dependence on the previous course of the match}

For the study of a possible  dependence on history we start by
analyzing the dependence on the order of the previous goals. As a
specific example we take all matches with the score 1:1 at half
time and check whether it makes a difference which team scored the
first goal of the match. If home team  scores the first goal one
obtains an average goal difference of 0.43 $\pm$ 0.07 for the
second half of the match. In the other case one obtains  0.46
$\pm$ 0.07. Within statistical uncertainty no difference can be
observed so that in this specific case the order of goals plays no
role. This motivates our choice to neglect the specific order of
the previous goals in what follows.

\begin{figure}[tb]
\centering\includegraphics[width=0.7\columnwidth]{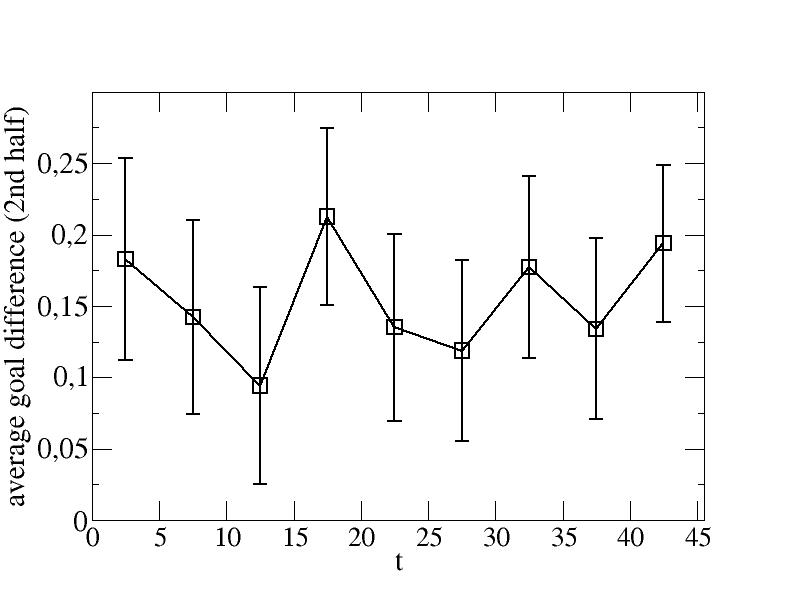}
\caption{The goal difference in the second half of the team leading by
1:0 at half time. Shown is the dependence on the
minute t when this single goal has been scored.} \label{fig3}
\end{figure}

Next we study a possible dependence on the exact time of a goal.
Following the general ideas of Ref.\cite{And08} we record the
outcome of the second half of a match under the condition that one
team leads 1:0 at half time. Under this condition the leading team
will score on average approx. 0.15 goals more than the opponent in
the second half. This observation just reflects the fact that the
team, leading 1:0, tends to be the favorite and thus more likely
will be also more successful in the second half. As seen in Fig.
\ref{fig3} and in agreement with the results of Ref.\cite{And08}
the outcome of the second half is fully uncorrelated to the time
of the first goal. Thus, a goal just before half time is no more
influential on the further course of the match than an earlier
goal. As already mentioned above this observation simplifies the
statistical description of a soccer match because it is not
important to keep into account the absolute values of the scoring
times.

\begin{figure}[tb]
\centering\includegraphics[width=0.7\columnwidth]{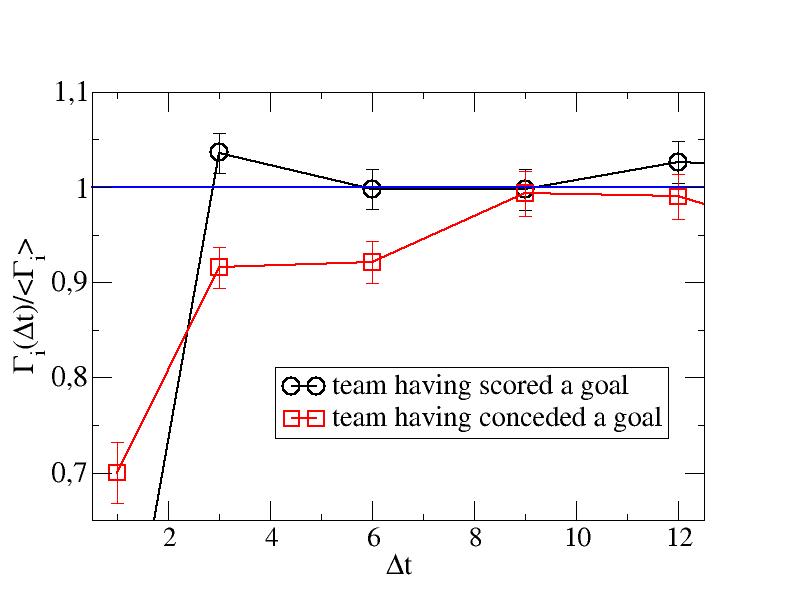}
\caption{The goal rate of a team if a time $\Delta t$ before it
has scored or conceded a goal, respectively. The data have been
normalized such that the average for $t > 15$ is unity.
Except for $\Delta t=1$ all data points result from an average over three minutes.} \label{fig4}
\end{figure}

In the next  step we study the time which passed since the last
goal. Does scoring a goal change the pattern of the soccer match
in the minutes after this goal? For this purpose we analyze the
goal rate $\Gamma(\Delta t)$ at time $t + \Delta t$ if at time $t$
a goal had been scored. In analogy to our discussion in Sect. II
we average over all instance  where this condition is fulfilled.
Here we distinguish whether the same team or the opponent had
scored that goal (expressed by the index $i$ in Fig.\ref{fig4}).
The data are normalized by $\langle \Gamma(\Delta t \gg 1) \rangle
$, i.e. the typical goal rate sufficiently far away from the goal
under consideration. In this way possible anomalies after scoring
a goal can be directly read off for small $\Delta t$. We start
with the team which has scored the goal at time $t$. Naturally, at
time $t+1$ its goal rate is trivially suppressed because (a) there
is a short break and (b) the opponent receives the ball. However,
very soon the goal rate has reached its long-time limit.  The data
point at time $t+3$ is slightly increased by approx. 3\%. However,
this minor increase may be related to statistical fluctuations.
Thus, apart from the trivial short-time effect, scoring a goal has
no significant impact for that team. Most interestingly, this is
not true for the opposing team. Here the initial goal rate is
suppressed by approx. 10\% and the equilibration roughly takes
nearly 10 minutes. Thus, after a goal this team is less active for
a couple of minutes. This more defensive attitude, however, does
not reduce the number of conceded goals. Again there is an anomaly
for minute $t+1$ because of the break after the goal.

In summary, apart from the minor short-time effects as displayed
in Fig.\ref{fig4} no memory effects are present. Stated
differently: playing soccer is to a large extent a Markovian
process, i.e. the action during the next minute does not depend on
the time-history of how the present score has been generated.

\subsection{Dependence on the present score}

\begin{figure}[tb]
\centering\includegraphics[width=0.7\columnwidth]{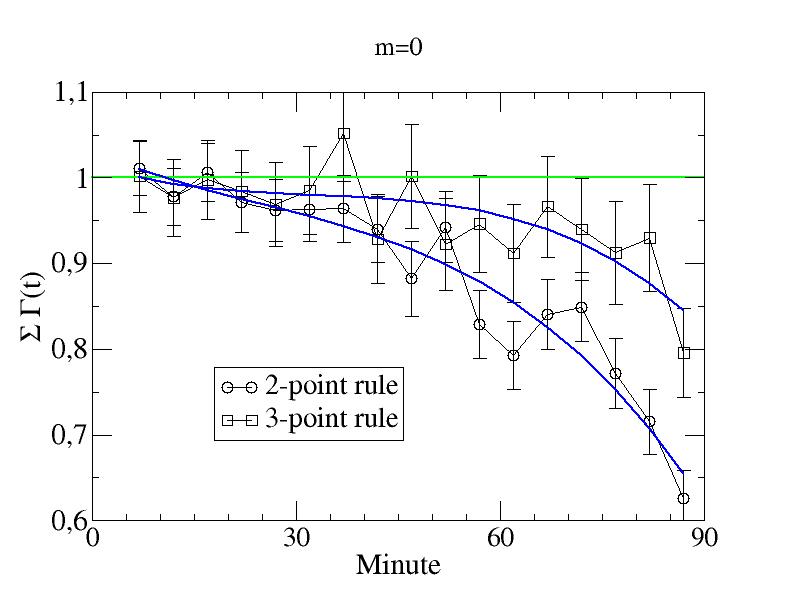}
\caption{The normalized number of goals in minute $t+1$ if in
minute $t$ the score is 0:0. Distinguished are the seasons with
the 2-point rule and with the 3-point rule, i.e. before and after
1995.} \label{fig7}
\end{figure}

In the remaining part of this work we analyse the question whether
the present score has any influence on the goal rate. We start by
analyzing the total goal rate, i.e. $\Sigma \Gamma \equiv \Gamma_h + \Gamma_a$  at time
$t$ when the score is 0:0.  Starting already from the
middle of the first half of the match the total goal rate tends to decrease.
During the last five minutes the rate is nearly 40 \% smaller than
the average rate. Actually this value was observed for the seasons
until 1994/95. Afterwards the 3-point rule has been introduced,
rendering the draw more unfavorable. Indeed the effect became less
relevant but is still significant (approx. 20\% reduction during
the last 5 minutes). Thus, the coziness of a draw is still present
in the heads of the soccer players.

\begin{figure}[tb]
\centering\includegraphics[width=0.7\columnwidth]{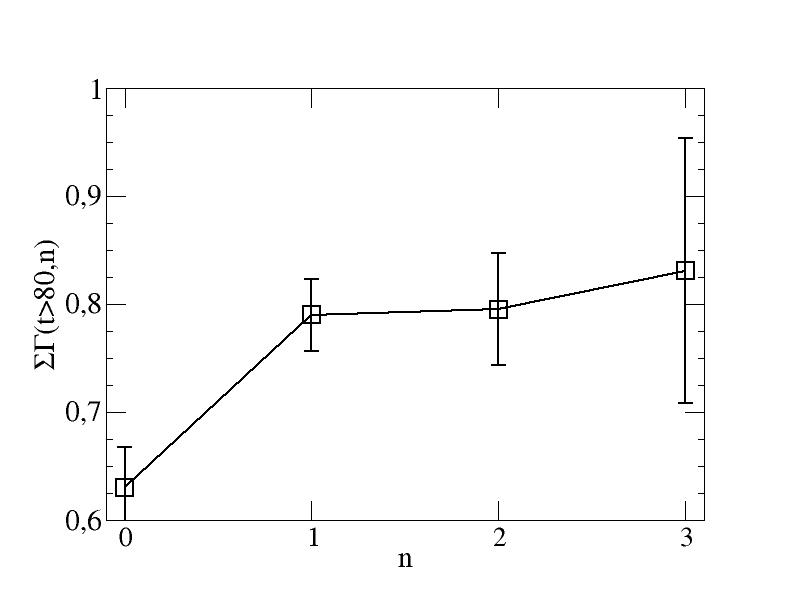}
\caption{The normalized number of goals in the last 10 minutes
under the condition that at minute $t=80$ the score is $n:n$.}
\label{fig8}
\end{figure}

Actually, this effect is particularly pronounced  in case of a
0:0. This can be seen from Fig.\ref{fig8} where we have plotted
the reduction in the total goal rate vs. the score (0:0,1:1,etc.)
during the last ten minutes. More specifically, we have calculated
 the goal rate at time $t$ if the
score is $n:n$ and then averaged this rate over $t \in [81,90]$. It
turns out that for all different types of draws the players seem
to be happy with the result, even in case of a 3:3 where already 6
goals had been scored. In summary, we have found the first example
where the match behavior significantly depends on the present
score.

Of course, as an immediate consequence the total goal rate in case
of a non-zero goal difference has to increase beyond unity
(approx. 1.07) because by definition the average over all matches
has to be unity.

\begin{figure}[tb]
\centering\includegraphics[width=0.7\columnwidth]{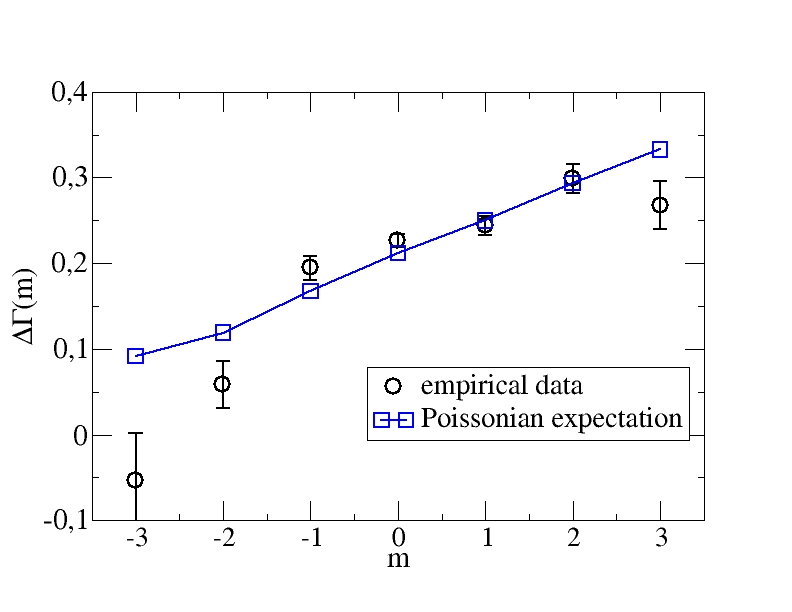}
\caption{The normalized difference of goal rates if the
goal difference is $m$  Included is the Poisson expectation.}
\label{fig5}
\end{figure}

More interesting than the sum of the goal rates is the difference
as expressed by $\Delta \Gamma \equiv \Gamma_h - \Gamma_a$. It
contains the information whether the balance between both teams in
a match is disturbed due to the present score. Specifically, we
consider $\Delta \Gamma(t,m)$, i.e. the dependence on time and
goal difference. In a first step we average $\Delta \Gamma(t,m)$
over all times with weighting factors which take into account the
relevance of a specific score at time $t$. The resulting
observable is denoted $\Delta \Gamma(m)$. One might speculate that
a team which is far ahead has an excellent day and very likely
will further increase its lead in the near future. Indeed, we find
in Fig.\ref{fig5} a monotonous increase of $\Delta \Gamma(m)$ with
the value of $m$. The fact that for all values of $m$ except for
$m=-3$ this difference is positive just represents the omnipresent
home advantage.

There are two extreme cases to explain this behavior. First, there
exist psychological effects which may render sportsmen more
successful in case of a lead. This effect would go beyond the
simple Poisson scenario and complicate the statistical description
of soccer matches. Second, it is a simple selection effect. Teams,
leading by three goals simply tend to be the better team and
therefore also score more goals in the future. In contrast to the
first scenario the future success is thus already determined
before the match and just reflects the different team strengths.

We can check whether the fitness variation among different teams
can fully explain the observed behavior. For this purpose we
additionally estimate the probability that the home or the away
team will score a goal in minute $t$ for the case of simple
Poisson statistics. According to Sect.2 we simply estimate
$\Lambda_h - \Lambda_a$ for the corresponding matches. In this way
we may take into account the above-mentioned selection effects.
These Poisson estimations are also included in Fig.\ref{fig5}.
One can clearly see that to a large extent the increase of $\Delta
\Gamma(m)$ with $m$ can be explained by the selection effect.
Thus, self-affirmative processes \cite{janke1,janke2,Heuer2} are not required to explain why
with a score of 2:0 the home team will be more successful in the
next minute as compared to a 0:2 situation. More generally, to a
first approximation the present score of the match does not
influence the relative success of both teams in the near future.

However, a more detailed analysis reveals some deviations from the
simple Poisson estimation. For $m=3$ there seems to be some
type of saturation mechanism which slightly reduces the tendency
to further increase the goal difference. Similarly, for $m=-2$ or
$m=-3$ the home team has a tendency to resign.

\begin{figure}[tb]
\centering\includegraphics[width=0.7\columnwidth]{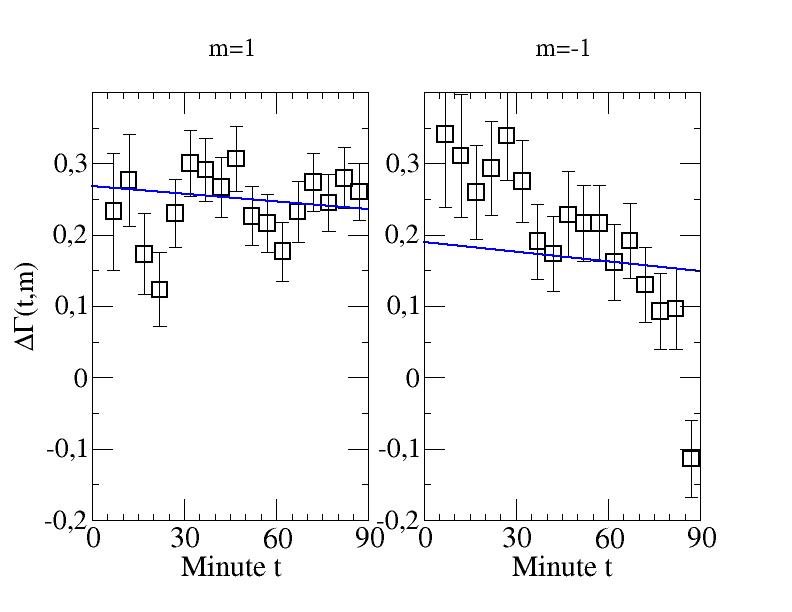}
\caption{The normalized difference of goal rates in minute $t$ under the
condition that at minute $t$ either the home team (left, $m=1$) or
the away team (right, $m=-1$) leads by one goal. Included is the
Poisson expectation.} \label{fig10}
\end{figure}

In analogy to Fig.\ref{fig7} we now present the time-resolved
rates. We compare the cases where either the home team or the away
team leads by one goal in minute $t$. We start the discussion with
$m=1$. Evidently, it is more likely in the next minute to
increases the one-goal lead rather than to reach a draw. This is
mainly an effect of the home advantage. To first approximation the
tendency towards an increasing lead is independent of the time of
the match. Furthermore, its value is fully consistent with the
Poisson expectation.  We may conclude that in case of one-goal
lead of the home team the match behavior follows a simple
Poisson behavior.

A closer inspection of $\Lambda_h - \Lambda_a$ shows that this
value somewhat decreases with increasing time. This has a simple
interpretation. If the first goal of the home team is already
scored after 10 minutes it is more likely that the home team is by
far the stronger team. Due to the highly random nature of a soccer
match this effect is weak and the discussion of Fig.\ref{fig7} is
not influenced by the weak time-dependence of $\Lambda_h - \Lambda_a$.

Most interestingly, the situation is very different if the away
team leads by one goal. One may distinguish three time regimes. In
case of an early lead (t $<$ 35) of the away team the home team is
more successful than expected from the Poisson expectation
to equal the score. The increase of
$\Delta \Gamma$ as compared to the Poisson expectation is as
large as 50\%. For t $>$ 35 the course of the match behaves as
expected from the statistical behavior. However, for t $>85$  a
dramatic change is observed. Suddenly it becomes even more likely
that the away team scores the second goal as compared to a draw.

\begin{figure}[tb]
\centering\includegraphics[width=0.7\columnwidth]{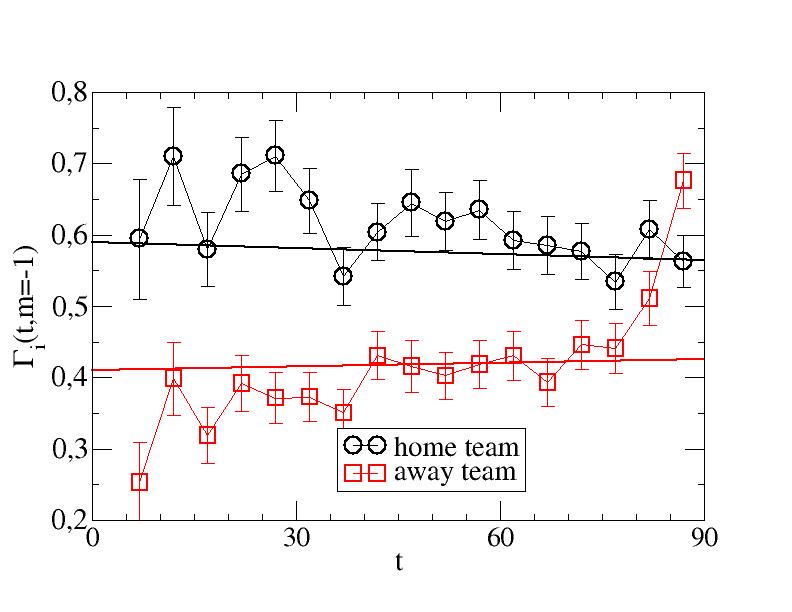}
\caption{The normalized number of goals in minute $t$ of the
home and away team, respectively, under the condition that
the away team leads by one goal. Included is the
Poisson expectation.} \label{fig13}
\end{figure}

In order to clarify the strong decay of $\Delta \Gamma$ during the
last minutes of a match we have individually determined $\Gamma_h$
and $\Gamma_a$; see Fig. \ref{fig13}. One can clearly see that the
anomalies at the end of the match are exclusively related to
$\Gamma_a$, i.e. the offensive of the away team and the defensive
of the home team. During the last 5--10 minutes the defensive of
the home team becomes much weaker. A straightforward
interpretation of this observation is the strengthening of the
offensive efforts at the expense of the defensive strength.
Unfortunately, on average these attempts are in vain because the
only effect is a larger number of conceded goals. In the most
extreme variant of this endeavor even the goal keeper starts to
support the own strikers. In any event, this behavior contributes
to the increase of $\gamma_{tot}(t)$ in the last minutes of the match as
discussed in Fig.\ref{fig1}.

The deviations from simple Poisson behavior for short times is
both related to an increase of $\Gamma_h$ and a decrease of
$\Gamma_a$. Obviously, in the first half of the match the home
team is still able in a focused and successful manner to intensify
its effort to reach a draw with an improvement in the offensive
and defensive part.

\begin{figure}[tb]
\centering\includegraphics[width=0.7\columnwidth]{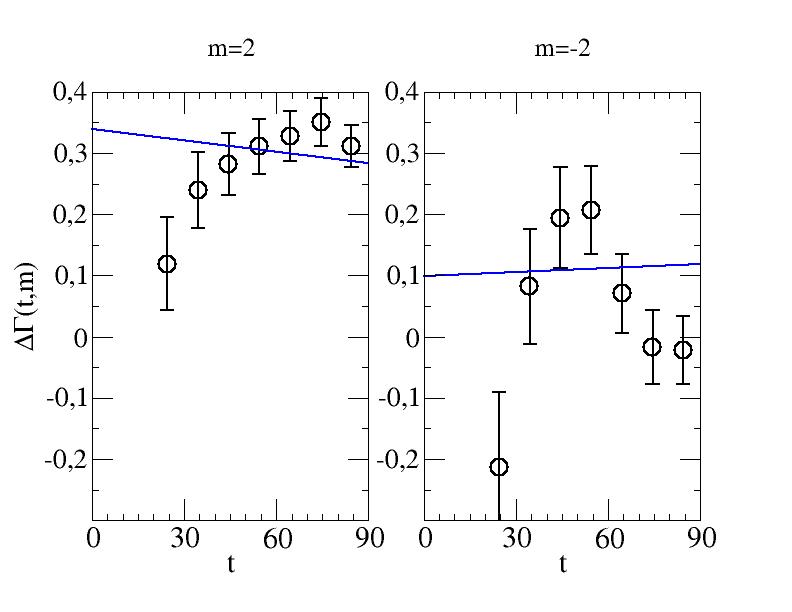}
\caption{The normalized goal difference in minute $t$ under the
condition that at minute $t$ either the home team (left, $m=2$) or
the away team (right, $m=-2$) leads by two goals. Included is the
Poisson expectation.} \label{fig102}
\end{figure}

We have repeated this analysis for $m=2$ and $m=-2$.  Here some
additional effects emerge. If the home team leads by two goals
very early in the match (around minute 20 to 30), the superiority
is smaller than expected from the Poisson expectation. Starting
from minute 40 a similar behavior is observed as for $m=1$. In the
opposite case $m=-2$ one observes that an early two-goal lag has
dramatic consequences on the performance of the home team. $\Delta
\Gamma$ is strongly reduced so that it is even more likely that
the away team scores the next goal. Note that the Poisson
expectation would still predict a small but significant home
advantage.  Only between minutes 40 and 60 the home team
successfully attemps to reduce a two-goal lag. Starting from
minute 60 these attemps start to be in vain.

\section{Discussion and Summary}

Based on  our systematic approach to identify deviations either
from the the Poisson expectation and/or from a strict Markovian
behavior we have obtained several key effects to characterize
complexities of soccer matches. (1) After a goal the opponent is
less successful to score a goal during the next minutes. This
invalidates a strict Markovian picture of soccer matches. This
effect, albeit significant, is relatively small (10\%). (2) In
case of a draw the total goal rate becomes smaller. Thus the goal
rates have to be adjusted in dependence of time and score. This is
a strong deviation from the Poisson expectation. (3) In case of a
lead of the away team dramatic deviations from the Poisson
expectation are observed during the last 5--10 minutes of the
match. This effect reflects inefficient defensive behavior of the home team.
The latter point indicates a dramatic difference of the behavior
of home and away teams which goes beyond the mere home advantage.
It signals strong psychological and/or tactical differences for
home as compared to away teams. Since the offensive efficiency does not become worse the present result does not imply that the home team gives up, at least in case of a one-goal lead by the away team. (4) If, however, the lead of the
away team occurs in the middle of the match there are indications
of an improved efficiency of the home team to equalize.

With respect to (1) it is interesting to refer to recent work on
scoring events in basketball. It has successfully been described
in terms of a biased continuous time random walk \cite{Gabel}.
Ideally the time difference between successive scoring events
should follow an exponential distribution. In practice already 20
seconds after a score the actual data follow very well this
theoretical expectation. In contrast, in the field of tennis statistics significant deviations from purely statistical behavior have been observed by Magnus and Klaassen \cite{Magnus}.
For example, after a break point it is more likely to win the next service game. Interestingly, this effect is more pronounced in matches between non-seeded players. This indicates that with increasing quality of the players the impact of previous effects become smaller,i.e. the match follows more a Markovian behaviour.

Our results also allow  us to find an answer to our initial
question about the origin of the large number of draws. It is the
persistence of a draw, i.e. (2), rather than the ability of a
team, trailing by one goal, to score an additional goal as
expressed by (3). Actually, (3) would rather decrease the number
of draws because the probability that a 0:1 transforms in a 0:2
during the last minutes is significantly larger than expected.

However, we should also stress that at least during the first  80
minutes most observables behave according to the simple Poisson
expectation as expressed by Eq. \ref{eqsimp}. This observation may
be used to discuss an important general question. Does the
empirical observation of a nearly Poisson-type goal distribution
imply that the process of scoring goals is indeed characterized by
some fixed rates? Alternatively one might postulate that good
teams try to achieve a safe lead and then just start to manage the
lead. In this scenario our in-match analysis should have detected
much larger deviations from Poisson behavior. For example one
might have guessed that $\Delta \Gamma(m=2)$ is much smaller than
expected from the Poisson scenario. Since this is not observed, the teams typically do not change their match behavior. Differences along this line
just start to (slightly) occur for $m=3$.

In summary, we may conclude that the concept of score-insensitive goal rates as opposed
to score-dependent match behavior is a very good approximation of a soccer match, at least
after averaging over the corresponding subset of matches as done in this work.
This naturally explains the previous observation \cite{Mah82, Lee97, Rue00, Heuer1} that
the goal distribution, after taking into account the different team strengths,
follows very nicely a Poisson distribution. This conclusion has an interesting
 consequence. A match of a good
team and bad team may have a priori goal expectations of 2 and 1,
respectively. A specific Poisson realization may, e.g.,  lead
to a 3:0 or (more unlikely) to a 1:3 result. In both realizations
the quality of the good team and that of the bad team are identical
because the final result is just a matter of mere luck (in analogy
to the presence or absence of the  decay of a radioactive nucleus
during a fixed time interval). In practice, one might expect that
in the first case media stress the successful play of the favorite
whereas in the second case the same team would be strongly
criticized. This reaction would neglected the random aspects,
inherent in any Poisson realization and just show that an objective
assessment of random aspects is very difficult.

It may be interesting in future work to check whether, e.g., the subset of good teams is less sensitive to negative effects (having just conceded a goal, leading behind at the end of the match). The present results may then serve as a detailed basis for the identification of possible strenth-dependent effects.

\section{Acknowledgement}

We acknowledge very helpful discussions with Bernd Strauss and
Dennis Riedl about this work.


\end{document}